\documentclass{elsart}

\usepackage{amsmath}
\usepackage{amssymb}
\usepackage{graphicx}
\usepackage{dcolumn}

\begin{document}

\begin{frontmatter}

\title{Superdiffusive Conduction: AC Conductivity with Correlated Noise}

\author[bra]{Fernando A.\ Oliveira}
\author[bra]{Rafael Morgado}
\author[nor]{Alex Hansen}
\author[spa]{J.\ M.\ Rubi}
\address[bra]
{Instituto de F{\'\i}sica and International Center for Condensed Matter Physics, Universidade de Bras{\'{\i}}lia, CP 04513,
70919--970, Bras{\'{\i}}lia-DF, Brazil.}
\address[nor]{Department of Physics, Norwegian University of Science and Technology, NO--7491 Trondheim, Norway.}
\address[spa]{Departamento de F{\'\i}sica Fondamental, Universitat de Barcelona, Diagonal 647, ES--08028 Barcelona, Spain.}





\begin{abstract}
We present evidence of the existence of a superdiffusive regime in
systems with correlated disorder for which localization is suppressed.
An expression for anomalous electrical conductivity at
low frequencies is found by using a generalized Langevin equation
whose memory function accounts for the interactions between the carriers.
New mechanisms inducing a  superdiffusive conductivity are discussed and
experimental possibilities for observing that phenomenon in nanotubes and
superlattices are presented.
\end{abstract}



\end{frontmatter}


One of the most fundamental quantities
in transport theory is electrical conductivity whose understanding has
permanently represented
a challenge for most of the last century. Discoveries of new aspects
of this property have always been accompanied by technological innovation.
Since finite conductivity is the consequence of the presence of one or
more scattering
mechanisms, most of the observed conductivity phenomena have always
been obtained for diffusive or subdiffusive propagation of the charge carriers.
The subdiffusive regime occurs in many different
disordered materials such as ion conducting materials or glassy plastics. 
Measurements of the AC conductivity as a function of the frequency reveal
the existence of universal behavior of this quantity whose origin seems to
be the only feature shared by those systems: the presence of
disorder \cite{Dyre}.

The presence of correlated disorder may introduce drastic changes
in the transport properties of a system by originating a superdiffusive
behavior of the charge carriers. This behavior can be observed e.g.\ in
the quantum Heisenberg chain \cite{Fio1} where one can find
the existence of both a superdiffusive regime emerging at weak correlations
and a ballistic regime resulting from strong correlations. A similar
result was obtained for a disordered chain \cite{Fio2}. It has been
proved as well that the Anderson model with long-range correlated
diagonal disorder displays a finite phase of extended states in the
middle of the band of allowed energies \cite{Fio0,Izrailev}. Moreover,
the suppression of Anderson localization has recently been confirmed
experimentally in semiconductor super-lattices with intentional correlated
disorder \cite{Bellani}. Quasi-ballistic and ballistic conduction
at room temperature has also recently been observed in carbon 
nanotubes \cite{Frank}.  

The aim of this Letter is to propose a general scenario to
describe the conduction mechanism in disordered materials valid for
both correlated and uncorrelated disorder. The result can be expressed by a
simple equation for the variation of the conductivity as function
of the frequency at low frequencies. We show that subdiffusive conductivity
increases with the frequency while superdiffusive conductivity decreases. The
result is general and independent from a particular microscopic
mechanism.

In the linear regime, transport properties can be determined through
the knowledge of  AC conductivity $\sigma(\omega)$. The first proposed
model for  AC conductivity, the Drude model, goes back more than a
century. In spite of great progress in the field, and the
discovery of many forms of universal behavior \cite{Dyre} no general model
accounting for the main characteristics of  complex disordered systems
has been set forth. Although a general relation for $\sigma(\omega)$ is not
available, some basic relations involving that quantity established long
ago still remain valid. The frequency-dependent diffusion coefficient
$D(\omega)$ is related to the conductivity through the expression
\begin{equation}
\sigma (\omega )=\frac{ne^{2}}{k_{B}T}D(\omega )\;,
\label{Cond}
\end{equation}
where, $e$ is carrier charge, $n$ the carrier density and $T$ is temperature 
and $k_B$ is the Boltzmann constant. 
The above equation is a form of the fluctuation-dissipation theorem. The 
frequency dependendent diffusion coefficient is defined as
\begin{equation}
D(\omega )=\int C_{v}(t)\exp (i\omega t)\ dt\;,
\label{D1}
\end{equation}
where $C_{v}(t)=\langle v(0)v(t)\rangle$ 
is the velocity correlation function of the 
carriers and the brackets
$\langle\cdots\rangle$ indicate ensemble average. 
Our aim here is to describe the interactions responsible for this behavior
using a generalized Langevin equation (GLE) which makes
the dependence more explicit and may display superdiffusion as well as
normal diffusion and subdiffusion.
Our starting point is that it does not matter which way $C_{v}(t)$ is determined 
once established, the diffusion and conductivity can be found. In 
general, complex materials present a non-Ohmic behavior in the conductivity, and
this effect is equivalent to the presence of memory, or a non-Markovian
behavior \cite{Kubo,Lee,Morgado,Costa}.

Conductivity of complex materials presents in
general a non-Ohmic behavior related to the presence of  memory effects
in the system. These effects have been successfully analyzed in many
Non-Markovian systems by means of the GLE. In our case, this equation is given 
by \cite{Kubo,Lee,Morgado,Costa} 
\begin{equation}
m\frac{dv(t)}{dt}=-m\int _{0}^{t}\Gamma (t-t')v(t')\ dt'+F(t)\;,
\label{1}
\end{equation}
where $\Gamma(t)$ is the memory kernel and
$F(t)$ is a stochastic noise subject to the conditions 
$\langle F(t)\rangle =0$, $\langle F(t)v(0)\rangle =0$ and
\begin{equation}
\langle F(t)F(0)\rangle =mk_{B}T\Gamma (t)\;,
\label{2}
\end{equation}
where $m$ is the effective mass of the carriers.
This last equation constitutes the formulation of the
fluctuation-dissipation theorem.

The memory function $\Gamma (t)$ can be directly defined as
\begin{equation}
\Gamma (t)=\int \rho _{n}(\omega )\cos (\omega t)\ d\omega\;,
\label{Gama}
\end{equation}
where $\rho _{n}(\omega )$ is the noise density of states or spectral
density. 
Our aim here is to use Eq.\ (\ref{1}) to analyze the existence of a
superdiffusion regime in conduction by computing the correlation
function and from there, the diffusion coefficient and the corresponding
conductivity according to formulae (\ref{Cond}).

An equation for $C_{v}(t)$ can readily be obtained from Eq.\ (\ref{1}) and 
the properties of the stochastic force as \cite{Morgado}
\begin{equation}
\frac{dC_{v}(t)}{dt}=-\int _{0}^{t}\Gamma (t-t')C_{v}(t')\ dt'\;.
\label{8.5}
\end{equation}
which can be solved, by means of the Laplace transform yields
\begin{equation}
\tilde{C_{v}}(z)=\frac{C_{v}(0)}{z+\tilde{\Gamma }(z)}\;,
\label{7}
\end{equation}
where $\tilde C_v(z)$ and $\tilde\Gamma(z)$ are the correlation function 
and memory kernel Laplace transforms. 
In view of Eq.\ (\ref{7}), Eq.\ (\ref{Cond}) can also be expressed as
\begin{equation}
\widetilde{\sigma }(z)=\frac{ne^{2}}{k_{B}T}\left(\frac{C_{v}(0)}{z+
\tilde{\Gamma }(z)}\right)\;,
\label{C2}
\end{equation}
in Laplace ($z$) rather than Fourier space ($\omega$).
This is our final form for the conductivity given in terms of the memory
function and the corrlation at the initial time. For systems
where the equipartition theorem holds, one expect that $C_{v}(0)=k_{B}T/m,$
and the conductivity becomes temperature independent. This feature has never 
been observed for
disordered materials; therefore, the only explanation of this behavior is
that equipartition is not fulfilled. It has been shown that for glassy
materials the fluctuation-dissipation theorem may fail \cite{PRR03} since the
system explores continuously metastable states never reaching true
thermal equilibrium. The same expression of the theorem holds in some
nonequilibrium situations if one replaces the temperature for an
effective temperature \cite{PRR03,Costa}. The failure of the 
fluctuation-dissipation relation has also been shown in systems undergoing 
ballistic motion \cite{Costa}.
In this situation the systems acquire an effective
temperature $T_{eff}\neq T$. This effective temperature is a signature
of metastability and has been observed in many instances
\cite{Costa}.  

The conductivity is in view of relation (\ref{Cond}) influenced
by the nature of the diffusion process through the value of the memory
function. Different diffusion regimes can be identified from the
asymptotic behavior of the mean square displacement
\begin{equation}
\lim _{t\rightarrow \infty }\langle x^{2}(t)\rangle\sim t^{\alpha }\;.
\label{3.5}
\end{equation}
This result follows from the GLE with the asymptotic expression
for the memory function 
$\lim _{z\rightarrow 0}\tilde{\Gamma }(z)=\tau ^{\nu -1}z^{\nu }$, where
$\tau$ is a relaxation time. The diffusion exponent $\alpha$ is connected
with $\nu$ through $\alpha=\nu+1$ \cite{Morgado}.
Consequently, subdiffusion occurs for $-1<\nu <0$,
normal diffusion for $\nu =0$, and superdiffusion for $0<\nu <1$.
For $\nu \geq 1$ the fluctuation-dissipation theorem fails \cite{Costa}.
Our result (\ref{C2}) explains the statement in Refs.\ \cite{Kimball,Dyre}
that at small frequencies $\widetilde\sigma(z)$ is an increasing function
of $z$ for subdiffusive propagation only.  
From Eq.\ (\ref{C2}) and the asymptotic value of the memory function 
given previously, one obtains
\begin{equation}
\frac{d\widetilde{\sigma }(z)}{dz}
\sim -\frac{1+\nu (z\tau )^{\nu -1}}{(z+\tilde{\Gamma}(z))^{2}}\;.
\label{widet}
\end{equation}
Consequently, for frequencies $z<\tau ^{-1}$, the derivative will
be positive for subdiffusion and normal diffusion, $-1<\nu \leq 0$,
and it will be negative for superdiffusion, $0<\nu <1$. 
The value for the second derivative of the conductivity
\begin{equation}
\lim _{z\rightarrow 0}\frac{d^{2}\widetilde{\sigma }(z)}{dz^{2}}
\thicksim -\frac{\nu (\nu -1)}{(\tau z)^{2}}\;
\label{limz}
\end{equation}
indicates that the curvature of the function $\widetilde\sigma(z)$ 
depends on $\nu$ and therefore predicts a different behavior of the 
conductivity in the diffusion regimes discussed. The frequency-dependent 
conductivity can be inferred from  Eq.\ (\ref{widet}) for small $z$, yielding
\begin{equation}
\frac{d\ln (\widetilde{\sigma }(z))}{d\ln (z)}\sim -\nu\;.
\label{fracd}
\end{equation}
Note that the previously obtained relations hold regardless of
the model used. By simply measuring the real part of the AC conductivity
$\sigma (\omega )$ as a function of the frequency $\omega $, we
can obtain information about the diffusive process and the density
of states, if the dimensionality of the system is known. If diffusion
takes place in a medium of dimensionality $d$, or in a fractal of
fractal dimension $d_{f}$, the diffusion coefficient scales as
\cite{Alexander}
\begin{equation}
D(\omega )\propto \omega ^{\alpha d_{f}-1}\:.
\label{dome}
\end{equation}
Theoretical studies have been performed in percolative clusters,
as well as experiments on material such as ionic conductive glass or amorphous
semiconductors, see \cite{Dyre,NYO94}, for reviews.
Those process have $\nu <0$, and are subdiffusive. Recent theoretical
\cite{Fio0,Fio1,Izrailev}, and experimental work \cite{Bellani,Frank},
may support our assumption of superdiffusive AC conductivity.

Between normal diffusion and ballistic propagation,
i.e., $0<\nu <1$, we may have many possibilities for superdiffusive
mechanism, some of which have been discussed for nanotubes and super-lattices.
The decrease of AC conductivity as a function of the frequency
is the main point we propose for experimental observation. Note
that the possibility of motion beyond ballistic, $\nu >1$, is very
rare. However, if they do exist, they violate the fluctuation-dissipation
theorem \cite{Costa}, and probably will not be described by linear response
theory.


Now we start to look for the possibility of superdiffusive or even
ballistic motion in some disorder-correlated system. Adame et al.\
\cite{Dominguez} investigated Bloc Oscillations in a one-dimensional lattice 
with long-range correlated diagonal disorder. For their system, the 
Schr{\"o}dinger equation for the Wannier amplitude is
\begin{equation}
i\frac{d\psi _{n}}{dt}=(\varepsilon _{n}-F)
\psi _{n}-\psi _{n+1}-\psi _{n-1}\;,
\label{idpsi}
\end{equation}
where $F$ is an external electrical field, and the stochastically  fluctuating
energies are
\begin{equation}
\varepsilon _{n}\propto \sum _{k=1}^{N/2}k^{-\beta /2}
\cos (\pi kn/N+\phi _{k})\;,
\label{energ}
\end{equation}
and $\phi _{k}$ are random phases in the interval {[}0,2$\pi ${]}.
For $\beta >2$, this model supports delocalized states with a clear
signature of Bloc oscillations. Recent studies \cite{Fio1} for the
quantum Heisenberg chain with $N$ sites, and exchange similar to
the energies above, present quite similar behavior. For example,
the spin wave function localized at the origin at $t=0$,
will present a superdiffusive motion with $\alpha =3/2$ for the weak
coupling, $0<\beta <1$. For strong coupling $\beta >1$, the diffusion
is ballistic. It has been proved as well that the Anderson model with
long-range correlated diagonal disorder displays a finite phase of
extended states in the middle of the band of allowed energies\cite{Fio0,Izrailev}.
Ballistic diffusion have been proposed as well for ratchet 
devices using a thermal broadband noise, with the elimination of the lower
modes \cite{Bao2}. Moreover, the supression of Anderson localization
was confirmed experimentally in semiconductor super-lattices with
intentional correlated disorder \cite{Bellani}. This behavior seems
to be universal insofar as a correlated disorder as the one above can be
defined. The correlation disorder is of course, the main scattering
mechanism which will be in charge of conduction. Some realistic
simulations would be helpful to find the best conditions to produce
that material. Despite the difficulties involved, experiments similar
to those by Bellani et al.\ \cite{Bellani}, may suggest that the 
stage is now set for superdiffusive conductors.


\end{document}